\documentclass[journal]{IEEEtran}
\usepackage{amsmath,amsfonts}
\usepackage{algorithmic}
\usepackage{array}
\usepackage[caption=false,font=normalsize,labelfont=sf,textfont=sf]{subfig}
\usepackage{textcomp}
\usepackage{hyperref}
\usepackage{stfloats}
\usepackage{url}
\usepackage{verbatim}
\usepackage{siunitx}
\usepackage{graphicx}
\def\BibTeX{{\rm B\kern-.05em{\sc i\kern-.025em b}\kern-.08em
    T\kern-.1667em\lower.7ex\hbox{E}\kern-.125emX}}
\usepackage{balance}
\usepackage{rotating} 
\usepackage{amsmath}
\usepackage{booktabs}
\usepackage{tikz}
\usetikzlibrary{arrows.meta, positioning}
\begin{document}
% \bstctlcite{Refs:use_etal}
\renewcommand{\thepage}{}
\title{Implementation of an Adaptive Transformer Accelerator for Accurate Outdoor Localization with Massive MIMO}
\author{Ilayda Yaman, Sijia Cheng, Ove Edfors, Liang Liu
\thanks{This work is funded by the Swedish Research Council, and ELLIIT (Excellence Center at Linköping-Lund in Information Technology), and partially supported by Ericsson AB and the Competence Center NextG2Com funded by the VINNOVA program for Advanced Digitalisation with grant number 2023-00541. \textit{Corresponding author: Ilayda Yaman.} % Manuscript created March 2025. 

The authors are with Department of Electrical and Information Technology, Lund University, Sweden (email:{\tt\small ilayda.yaman@eit.lth.se}).
}}

\maketitle

\begin{abstract}
We present a sparsity-aware FPGA implementation of an adaptive Transformer-based localization accelerator for 5G massive MIMO targeting sub-10\,ms real-time positioning. The architecture exploits propagation characteristics, where beam-delay channel representations exhibit sparsity, enabling a row-wise skipping mechanism that removes low-energy beam components with minimal control overhead. Transformer computations are mapped onto a heterogeneous vector processing engine with parallel processing elements and adder trees, using mixed input- and output-stationary dataflow execution for efficient matrix computation and reduced data movement. 
Environment-dependent processing is supported through a lightweight runtime model-switching mechanism, where temporally filtered outputs of a single-layer perceptron router enable selection between specialized models with reduced latency. Implemented on a Xilinx Zynq UltraScale+ FPGA and evaluated on real-world massive MIMO measurements, the design achieves up to 65\% row sparsity, yielding peak computational speedups of approximately 2x while limiting the average localization accuracy degradation to below 10\%, relative to the fixed-point baseline model. The accelerator attains below 1.15\,m localization accuracy across scenarios, with inference latency of 0.51-2.11\,ms and throughput of up to 1961 positions/s. These results demonstrate that propagation-aware sparsity, mixed dataflow execution, and efficient runtime model switching enable a scalable and low-latency hardware realization of adaptive Transformer-based localization for real-time 5G systems. 
\end{abstract}


\section{Introduction} 
Real-time localization systems require accurate position estimation under strict latency, throughput, and energy constraints, making efficient hardware implementation a fundamental challenge. In particular, massive multiple-input multiple-output (MIMO) systems generate computationally intensive localization workloads due to the high dimensionality of beam- and subcarrier-domain channel measurements acquired across large antenna arrays. 
At the same time, these measurements provide rich spatial signatures that can support meter-level localization, which is relevant for applications such as autonomous systems, intelligent transportation, and location-aware wireless services.

Machine learning methods such as convolutional neural networks (CNNs) have been used for massive MIMO based localization and implemented as application-specific hardware~\cite{Mohammad_mimo_pos, attari2025acceleratorassistedfloatingpointasipcommunication}. More recently, attention-based models, particularly Transformer architectures built upon self-attention mechanisms, have demonstrated strong performance in radio-based localization, enabling accurate positioning under diverse environmental conditions~\cite{tian2024attention, yaman2025adaptiveattentionbasedmodel5g, indoor_localization_attention_5g, EfficientLocNet}.  The attention mechanism~\cite{attentionisallyouneed} captures correlations between beam-domain channel responses, where pairwise similarity patterns provide valuable spatial information for user localization.

Achieving low-latency inference for massive MIMO localization remains challenging due to the computational complexity, memory requirements, and data movement associated with processing high-dimensional channel measurements in real time~\cite{Mohammad_mimo_pos}. 
Transformer-based models further intensify these challenges because self-attention introduces quadratic computational complexity with respect to sequence length, together with substantial memory traffic and synchronization overhead. In addition, radio-based localization workloads are highly dependent on propagation conditions, where line-of-sight (LoS), non-line-of-sight (NLoS), and multipath environments produce different channel structures and sparsity characteristics~\cite{wen2019}. These variations make it difficult for a single model to efficiently capture all scenarios without increasing model complexity and computational cost.

Specialized models can instead exploit the structured properties of radio signals within each propagation condition, enabling improved localization accuracy while reducing computational complexity through lower model size and increased input sparsity~\cite{yaman2025adaptiveattentionbasedmodel5g}.
A lightweight routing mechanism can then dynamically select the appropriate model based on the observed channel characteristics, enabling adaptive inference with reduced computational cost. Efficient hardware realization of such adaptive execution requires low-overhead model switching while maintaining real-time operation.

FPGA-based Transformer accelerators have largely targeted general-purpose vision and language workloads~\cite{fpga_dataflow_ftrans, desa, famous}, with techniques such as input-aware path selection, token pruning, and quantization used to reduce computational cost~\cite{swat, ayaka_transformer, sanger_sparse, moitra2024pivot, wang2020spatten}. However, these approaches are primarily developed for vision and language workloads and do not explicitly target the structured sparsity patterns inherent to massive MIMO beam-delay representations. 

Recent work has also explored sigmoid-based alternatives to softmax attention~\cite{softmax_sigmoid, softmax_sigmoid2}. Since sigmoid operates element-wise and avoids row-wise normalization, it is well suited to parallel, streaming, fixed-point, and lookup-table-based hardware implementations. 
When combined with biasing or normalization strategies, sigmoid attention can maintain competitive accuracy while significantly reducing hardware complexity. These properties make sigmoid attention particularly attractive for domain-specific accelerators targeting structured inputs such as beam–delay channel representations.

\begin{figure*}[tb!]
  \centering
  \includegraphics[width=1.0\linewidth]{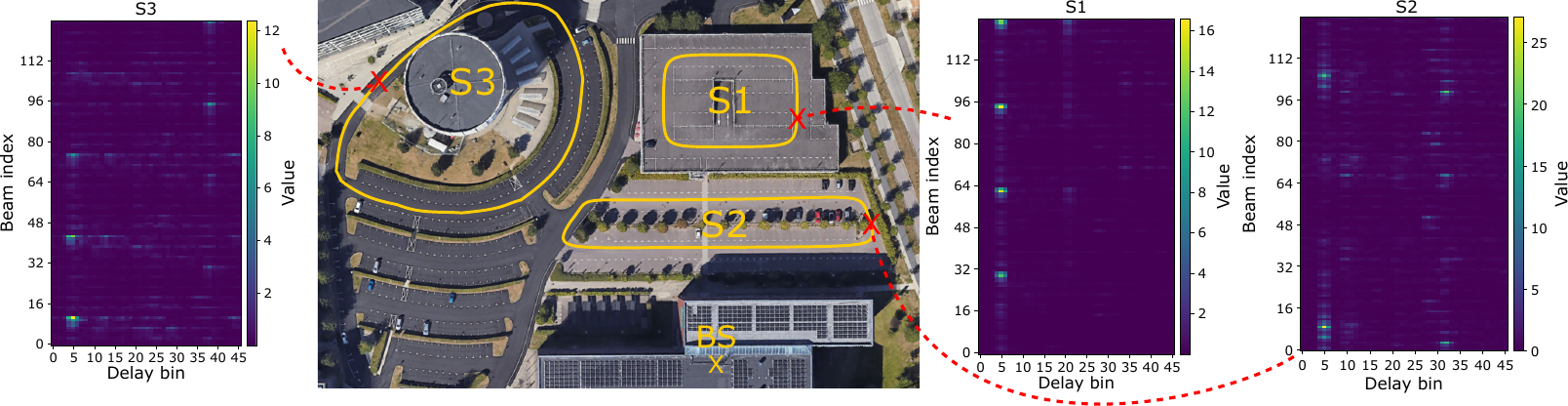}
  \caption{Bird-eye view of the measurement environment and trajectories labeled S1, S2, and S3. Example user positions for each scenario are marked with red cross and their corresponding fingerprints are shown.}
  \label{fig:measurement_env}
\end{figure*}

Building on these observations, this work presents a domain-aware (i.e., radio-localization-specific), adaptive, and reconfigurable Transformer accelerator tailored for radio-based localization.  
A lightweight single-layer perceptron (SLP) router enables switching between specialized models corresponding to different propagation scenarios with minimal control overhead. 
The architecture integrates sparsity-aware computation, sigmoid attention, and low-overhead model selection within a unified design. Symmetric fixed-point quantization is employed to enable integer-only inference, and multiple dataflow strategies are utilized to optimize Transformer computations. 
The resulting accelerator provides an energy-efficient solution for deploying adaptive localization models on resource-constrained hardware platforms. 
The design is validated using real-world channel measurements from a commercial 5G massive MIMO base station (BS) \footnote{Referred to as a gNodeB (gNB) in 5G systems.} and evaluated against the 3GPP Release~17 positioning requirements, which include low-latency targets on the order of $10$\,ms for selected use cases~\cite{3GPP_TS_22261,3GPP_TR_38857}. 

\section{System Model and Transformer-based Localization}

In radio systems, channel state information (CSI), describing the relationship between transmitted and received signals, can be used as fingerprints for machine learning-based localization. In this work, the user equipment (UE) transmits pilot signals, and the massive MIMO BS operating with 5G new radio (NR) orthogonal frequency-division multiplexing (OFDM), estimates the corresponding channel responses across beams and subcarriers. These measurements are processed by the proposed adaptive Transformer model to estimate spatial coordinates, such as the user position $(x, y)$. The predicted positions are compared against reference coordinates obtained from a UE-mounted GNSS receiver, and the localization accuracy is quantified using the mean Euclidean error (ME). Fig.~\ref{fig:measurement_env} shows a bird's-eye view of the measurement environment together with representative beam-delay profiles corresponding to selected positions.

\subsection{System Model}

The data is collected using a commercial massive MIMO BS and a single mobile UE mounted on a vehicle. The BS operates at a center frequency of $3.85$\,GHz with $100$\,MHz bandwidth and captures uplink sounding reference signal (SRS) measurements every $20$\,ms. Each snapshot contains channel estimates across $64$ beams and $273$ physical resource blocks (PRBs). To reduce dimensionality, adjacent PRBs are averaged and downsampled to $137$ subgroups, which are further reduced to $46$ subcarriers through interleaved sampling.

The BS antenna array consists of $32$ vertically and $32$ horizontally polarized antenna ports that form $64$ beams. The UE employs $4$ antenna ports, where two antenna pairs transmit SRS signals. For each snapshot $t$, the measured beam-space channel transfer functions from the two UE antenna pairs and both polarizations are represented by matrices $\mathbf{H}_{\mathrm{H1},t}$, $\mathbf{H}_{\mathrm{V1},t}$, $\mathbf{H}_{\mathrm{H2},t}$, and $\mathbf{H}_{\mathrm{V2},t}$. These matrices are stacked to form the combined channel tensor

\begin{equation}
\mathbf{H}_t \in \mathbb{C}^{128 \times 46}
=
[\mathbf{H}_{\mathrm{H1},t}^T,
 \mathbf{H}_{\mathrm{V1},t}^T,
 \mathbf{H}_{\mathrm{H2},t}^T,
 \mathbf{H}_{\mathrm{V2},t}^T]^T ,
\end{equation}
which represents the beam–frequency structure of the channel at a given time snapshot.

Details on the measurement campaign and preprocessing are provided in~\cite{tian2024attention, yaman2025adaptiveattentionbasedmodel5g, pjanic2025illuminating}. During preprocessing, a $46$-point Hann window is applied across the frequency dimension to suppress side lobes. An inverse FFT is then applied along the subcarrier dimension to obtain a beam-delay representation of the channel, denoted as $\mathbf{G}_t \in \mathbb{C}^{128 \times 46}$. Due to the potential difficulty of achieving stable phase information across measurements, the amplitude $|\mathbf{G}_t|$ is used as fingerprints for the localization model, yielding a tensor of size $128 \times 46$.

The UE follows three predefined trajectories, forming three propagation scenarios illustrated in Fig.~\ref{fig:measurement_env}, which also includes representative beam-delay matrices for each scenario. Scenario~S1 corresponds to a predominantly line-of-sight (LoS) environment located on the roof of a parking structure approximately $10$\,m above ground level. Scenario~S2 is located at ground level below the BS and mainly represents non-line-of-sight (NLoS) conditions. Scenario~S3 contains mixed propagation conditions with intermittent LoS and obstructions caused by nearby structures. The vehicle travels along each trajectory for five laps at a target average speed of $15$\,km/h while maintaining a fixed antenna orientation, relative to the vehicle, on the vehicle roof. Based on the reference GNSS measurements and Euclidean distances between samples, the spatial sampling density differs slightly between the scenarios. For the evaluation lap, S1 covers approximately $140$\,m over $4200$ snapshots (about $3$\,cm per step), S2 covers $245$\,m over $4850$ snapshots (approximately $5$\,cm per step), and S3 covers $265$\,m over $5000$ snapshots (approximately $5$\,cm per step). 

\subsection{Transformer-based Localization}

\begin{figure}[tb!]
  \centering
  \includegraphics[width=1.0\linewidth]{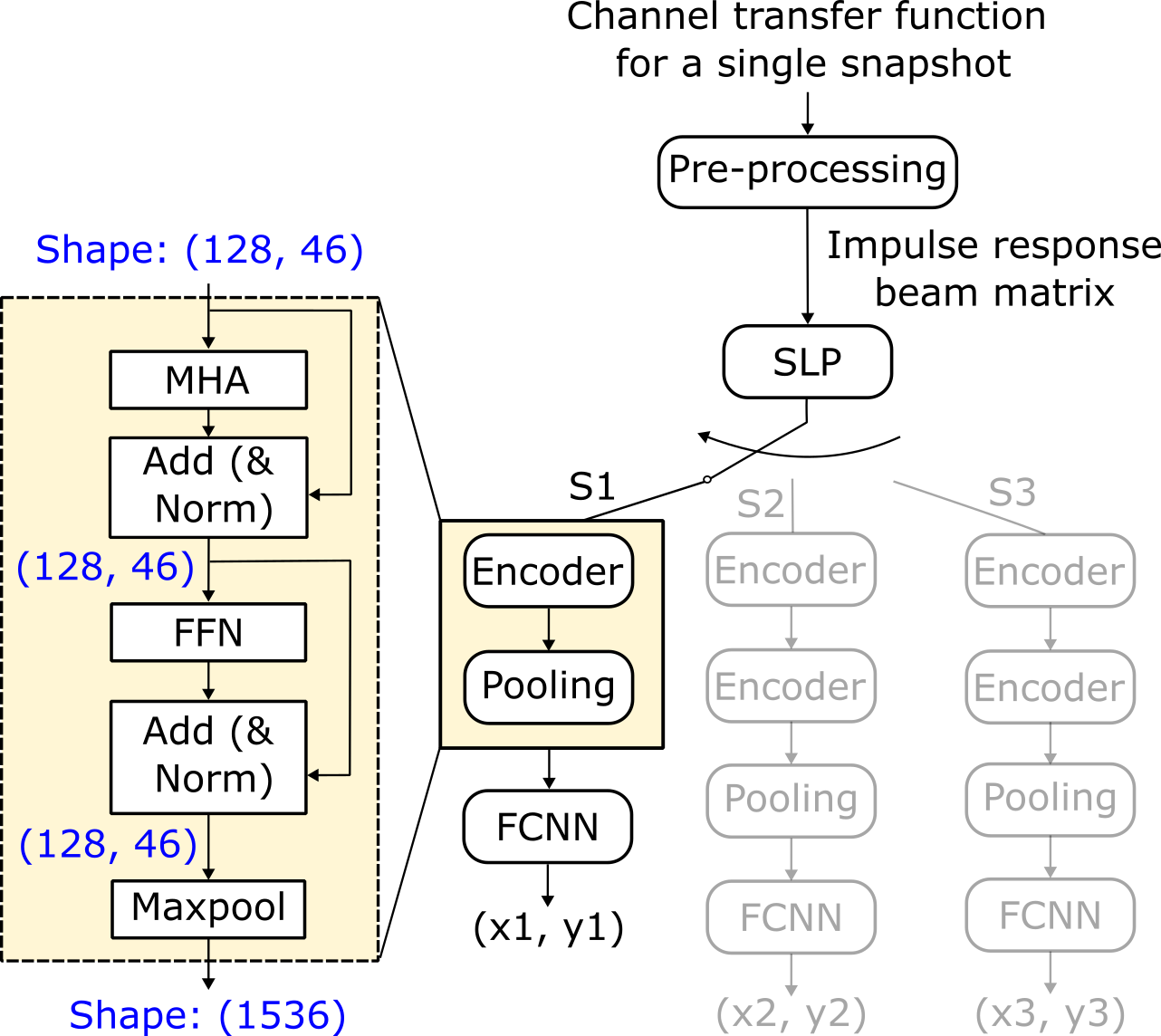}
  \caption{Overall view of the adaptive localization system, featuring the SLP router and three specialized Transformer-based models. Input data dimensions at each sub-layer are indicated in blue.}
  \label{fig:high_level_diagram}
\end{figure}

The adaptive localization system, shown in Fig.~\ref{fig:high_level_diagram}, analyzes input characteristics and environmental conditions to dynamically select between three specialized Transformer-based models~\cite{yaman2025adaptiveattentionbasedmodel5g}. The specialized models are separately trained for the propagation scenarios associated with the trajectories labeled S1, S2, and S3 in Fig.~\ref{fig:measurement_env}. To balance localization accuracy and computational cost, the specialized models employ different architectural depths and trainable parameters and a SLP acts as a lightweight router to select the most suitable specialized model for each channel snapshot, enabling reliable and efficient localization while optimizing the trade-off between localization accuracy and processing latency.

\subsubsection{Single-Layer Perceptron (SLP)}
An SLP is employed as a lightweight routing module to select between multiple specialized models. Instead of using the full beam–delay representation, a single delay bin is extracted from the input, yielding a feature vector $\mathbf{x} \in \mathbb{R}^{128}$ corresponding to a randomly selected fixed delay in $|\mathbf{G}_t|$.
The SLP produces a vector of class logits:
\begin{equation}
    \mathbf{y} = \mathbf{W}\mathbf{x} + \mathbf{b},
\end{equation}
where $\mathbf{W} \in \mathbb{R}^{3 \times 128}$ and $\mathbf{b} \in \mathbb{R}^{3}$ are learned parameters, and each element of $\mathbf{y}$ corresponds to a propagation scenario class. 

During training, the logits are passed to a cross-entropy loss, which internally applies a softmax operation to obtain class probabilities and encourages separation between the correct and competing classes. At inference, the predicted class is obtained by selecting the index of the maximum logit, which determines the specific model to use for the given input. Due to its linear structure and reduced input dimensionality, the SLP introduces minimal computational and hardware overhead, making it well suited for real-time adaptive model selection.

\subsubsection{Multi-Head Attention (MHA)}
The input to the localization model for a single snapshot at time $t$ is the amplitude of the preprocessed beam-delay channel matrix described in Section II-A, defined as $\mathbf{X} = |\mathbf{G}_t| \in \mathbb{R}^{128 \times 46}$, where $n=128$ denotes the sequence length corresponding to the number of beam tokens, and $d=46$ corresponds to the delay-bin features associated with each beam. The model employs a self-attention mechanism in which the query ($\mathbf{Q}$), key ($\mathbf{K}$), and value ($\mathbf{V}$) matrices are derived from the same input $\mathbf{X}$ through linear projections:
\begin{equation}
    \mathbf{Q} = \mathbf{X}\mathbf{W}_q, \quad
    \mathbf{K} = \mathbf{X}\mathbf{W}_k, \quad
    \mathbf{V} = \mathbf{X}\mathbf{W}_v,
\end{equation}
where $\mathbf{W}_q, \mathbf{W}_k, \mathbf{W}_v \in \mathbb{R}^{d \times d}$  are learned weight matrices, yielding $\mathbf{Q}, \mathbf{K}, \mathbf{V} \in \mathbb{R}^{n \times d}$. In multi-head attention, these matrices are evenly split along the feature dimension across $h$ heads, such that each head operates on a subspace of dimension $d_k = d_v = d/h$. In this work, $h=2$.

The attention scores for each head are computed using scaled dot-product attention with an additional learnable scaling parameter $\gamma$:
\begin{equation}
    \mathbf{S}^i = \gamma \frac{\mathbf{Q}^i{\mathbf{K}^i}^T}{\sqrt{d_k}},
    \label{eq:before_attention_scores}
\end{equation}
where $\mathbf{Q}^i, \mathbf{K}^i \in \mathbb{R}^{n \times d_k}$ are the $i$-th partitions of $\mathbf{Q}$ and $\mathbf{K}$, and $\sqrt{d_k}$ prevents the dot-product magnitudes from growing large, thereby stabilizing training. 
The learnable parameter $\gamma$ modulates the magnitude of the attention logits, allowing the model to control the sharpness of the resulting attention distribution~\cite{ram2025learningfocusfocalattention}. 

The output of the $i$-th head is then:
\begin{equation}
    \text{head}_i = \phi(\mathbf{S}^i)\mathbf{V}^i,
    \label{eq:attention_scores}
\end{equation}
where $\mathbf{V}^i \in \mathbb{R}^{n \times d_k}$ is the $i$-th partition 
of $\mathbf{V}$, and $\phi(\cdot)$ is the activation function applied to the attention scores. The activation generates non-negative weights used to compute a weighted sum of the value vectors. For softmax, the weights are normalized across each row to sum to one, whereas sigmoid produces element-wise values in the range $[0,1]$ without normalization. In this work, both variants are analyzed.

The outputs of all heads are concatenated and linearly projected:
\begin{equation}
    \text{MHA}(\mathbf{X}) =
    \text{Concat}(\text{head}_1,\dots,\text{head}_h)\mathbf{W}_o,
\end{equation}
where $\mathbf{W}_o$ is a learned output projection matrix. The resulting representation is then combined with the input through a residual connection.

\subsubsection{Feed-Forward Network (FFN)}

Following each self-attention layer, a position-wise FFN is applied independently to each token. The FFN consists of two linear transformations with a non-linear activation function, the rectified linear unit (ReLU), in between:
\begin{equation}
    \text{FFN}(\mathbf{X}) =\text{ReLU}(\mathbf{X} \mathbf{W}_1  + \mathbf{b}_1) \mathbf{W}_2 + \mathbf{b}_2,
\end{equation}
where $\mathbf{W}_1 \in \mathbb{R}^{d \times d_{\text{ff}}}$ and $\mathbf{W}_2 \in \mathbb{R}^{d_{\text{ff}} \times d}$, with $d_{\text{ff}} = 64$ denoting the hidden dimension of the FFN.

\begin{figure*}[tb!]
  \centering
  \includegraphics[width=1.0\linewidth]{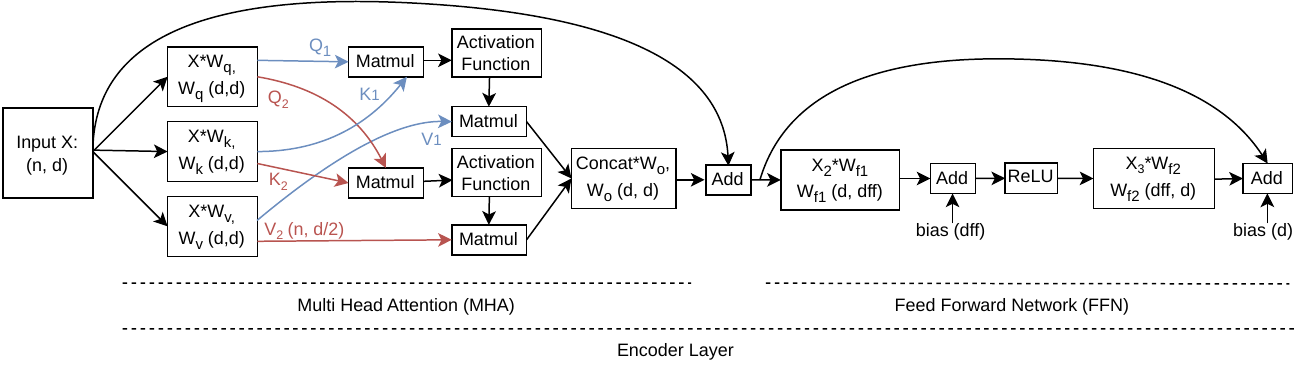}
  \caption{Block diagram of a single encoder layer with two attention heads, showing the dimensions of inputs, weights, and biases.}
  \label{fig:complete_attention}
\end{figure*}

When MHA and FFN are stacked together, as illustrated in Fig.~\ref{fig:complete_attention}, they form one encoder layer.

\subsubsection{Fully Connected Neural Network (FCNN)}

Following the encoder layers, a max-pooling operation is applied along the feature dimension, resulting in a reduced representation of size $(n \times \frac{d+p}{k})$, where $p$ denotes zero-padding applied to ensure alignment with the pooling factor $k$. The pooled tensor is then flattened and passed through an FCNN consisting of two linear layers:
\begin{equation}
    \mathbf{y} = \sigma(\mathbf{X} \mathbf{W}_1+ \mathbf{b}_1) \mathbf{W}_2 + \mathbf{b}_2,
\end{equation}
where $\mathbf{W}_1 \in \mathbb{R}^{(n(d+p)/k) \times d_h}$ and $\mathbf{W}_2 \in \mathbb{R}^{d_h \times d_{\mathrm{out}}}$ are learned weight matrices, $d_h$ denotes the hidden dimension, and $d_{\mathrm{out}}=2$ corresponds to the predicted $(x,y)$ coordinates. The activation function $\sigma(\cdot)$ is implemented as a leaky ReLU with a negative slope of $0.3$.

\section{Input Sparsity Exploration}

As seen in Fig.~\ref{fig:measurement_env}, in the beam domain, a small subset of beamforming directions carries significant channel energy due to the limited number of dominant propagation paths between the UE and BS. Similarly, in the delay domain, only a few multipath components contribute significant energy, as some reflections arrive with distinct delays while the remaining components are weak.
This results in a coarse-grained sparsity pattern in the beam–delay representation. To exploit this property, a two-stage procedure is applied: element-wise thresholding to suppress low-magnitude components, followed by row-wise sparsity detection, where each row is evaluated based on a zero-count criterion to identify beams that can be skipped.

\paragraph{Element-wise Thresholding}
Many values in the beam-delay representation have small magnitudes and contribute negligibly to the final localization output. To eliminate these insignificant values, an element threshold $T$ is applied. Specifically, any matrix element satisfying

\begin{equation}
x_{ij}  < T
\end{equation}
is treated as zero for the purpose of sparsity detection. Different values of $T$ (e.g., $10^{-6}$, $10^{-5}$, $10^{-4}$) are evaluated to determine which elements can be safely ignored without significantly degrading localization accuracy. Since the distribution of channel magnitudes varies across propagation environments, the optimal threshold may differ depending on the scenario (e.g., LoS or NLoS). In addition to element-wise thresholding, the aggregate power of each row is analyzed to characterize the contribution of different beams. 

\paragraph{Row-wise Sparsity Detection}

After element-wise thresholding, the number of zero elements within each row is counted. Let $Z_i$ denote the number of zero elements in row $i$. If the number of zeros exceeds a predefined row sparsity threshold $T_r$, the entire row is masked out:

\begin{equation}
Z_i > T_r \quad \Rightarrow \quad \text{row } i \text{ is skipped}.
\end{equation}
This mechanism allows rows that contain predominantly near-zero values to be removed from the computation. The choice of $T_r$ determines how aggressively rows are skipped and can also vary depending on the propagation scenario, since different environments produce different sparsity patterns.
When a row is masked, all multiply-accumulate operations associated with that row are skipped in the subsequent matrix multiplications. 

\begin{figure*}[tb!]
  \centering
  \includegraphics[width=1.0\linewidth]{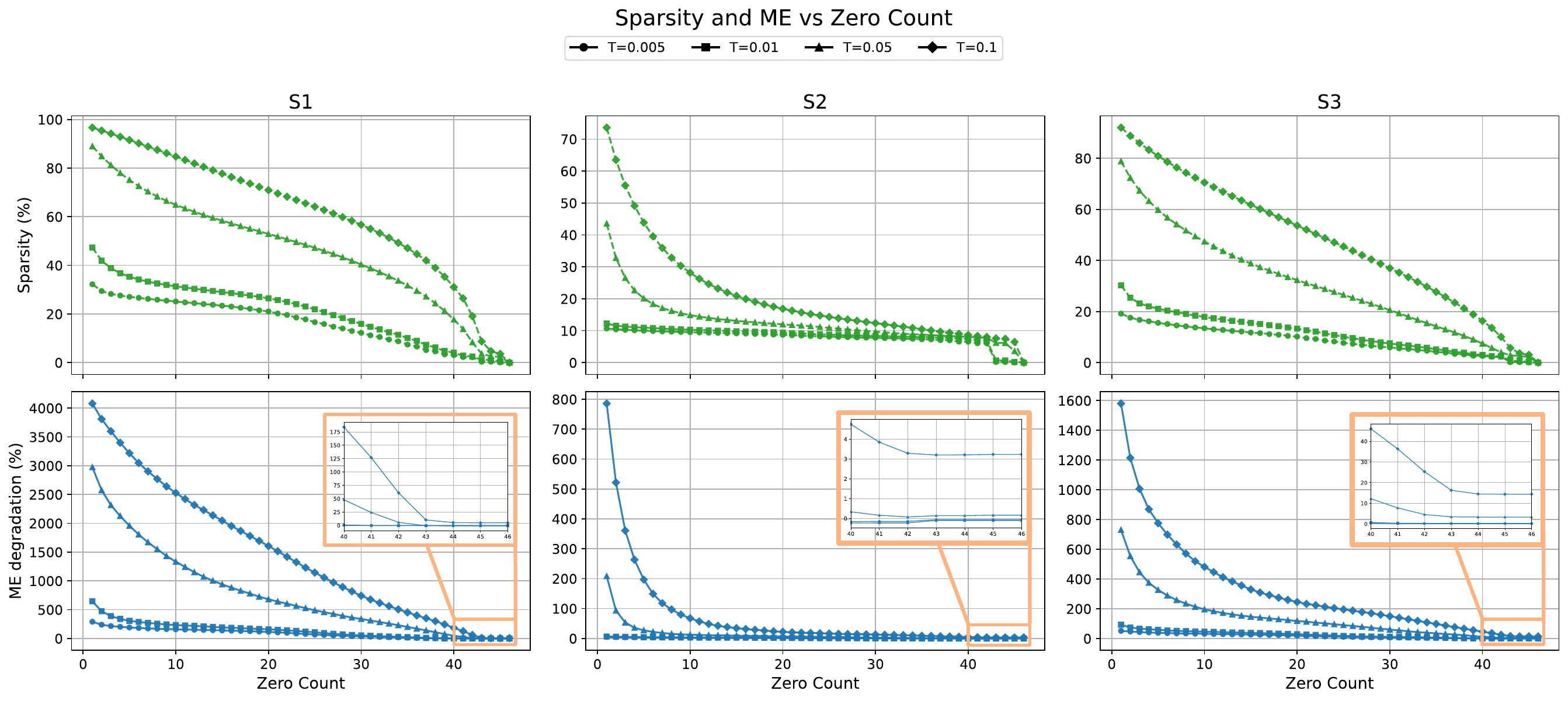}
  \caption{The effect of threshold value (T) and zero count in a row ($T_r$) on ME and the total number of rows calculated on S1, S2 and S3.}
  \label{fig:sparsity_t_zc_s1}
\end{figure*}

To quantify the amount of row-wise sparsity present in the data, we analyze the distribution of near-zero elements in the input matrices. For each snapshot the $(128,46)$ input matrix is examined row-wise, and the number of elements below a threshold value $T$ is counted. 
Fig.~\ref{fig:sparsity_t_zc_s1} illustrates the relationship between the selected threshold values $T$, the number of near-zero elements per row, and the resulting ME.

A full parameter sweep was conducted over the zero-count threshold $(0,1,\dots,46)$ and threshold $(0.001$ to $0.1)$ to identify suitable sparsification settings for each data type. The selection criterion aims to maximize row-wise sparsity while constraining the degradation in ME to within $10\%$ relative to the baseline. For each data type (S1, S2, S3), the results in Table~\ref{tab:summary_input_sparsity} correspond to the highest sparsity level that satisfies this error constraint. 

\begin{table}[htbp!]
\caption{Sparsity comparison.}
\begin{center}
\begin{tabular}{|l|c|c|c|}
\hline
\textbf{Data Type} & \textbf{S1} & \textbf{S2}& \textbf{S3}\\ \hline
Threshold & 0.039 & 0.014 & 0.003  \\ \hline
Zero Count & 41 & 1 & 24 \\ \hline
Element Sparsity (\%) & 43.6 & 14.2 & 9.25 \\ \hline
Row Sparsity (\%) & 10.1 & 14.0 & 6.99 \\ \hline
ME degradation (\%) & 9.5 & 9.6 & 9.9  \\ \hline
\end{tabular}
\label{tab:summary_input_sparsity}
  \end{center}
\end{table}    

\section{Activation functions} 
\label{act_func}

The activation function in the attention mechanism is a key determinant of both computational efficiency and hardware complexity in Transformer accelerators. In particular, the softmax operation introduces global data dependencies due to the required maximum, exponential, and normalization steps, limiting parallelism and increasing implementation cost. Prior work such as DESA~\cite{desa} identifies softmax as a major bottleneck in the attention pipeline in terms of both latency and resource utilization. These challenges arise from row-wise reductions and normalization operations, which require intermediate buffering and restrict streaming execution.

To address these limitations, alternative activation functions based on sigmoid have been proposed~\cite{softmax_sigmoid, softmax_sigmoid2}. These approaches aim to reduce global dependencies and improve hardware efficiency. ReLU-based attention has also been explored~\cite{studyrelusoftmaxtransformer, replacingsoftmaxreluvision}, but preliminary experiments showed degraded localization performance in this application and it is therefore not considered further for this architecture.

\subsection{Softmax-Based Self-Attention}

The softmax function converts a vector of logits $\mathbf{x}=[x_1,\dots,x_n]$ into normalized weights
\begin{equation}
\alpha_i = \frac{e^{x_i}}{\sum_{j=1}^{n} e^{x_j}} .
\end{equation}

In the attention mechanism, the logits correspond to each row of the attention score matrix $\mathbf{S}$. Specifically, for row $i$, we have $x_j = S_{ij}$, yielding the attention weights
\begin{equation}
A_{ij} =
\frac{e^{S_{ij}}}{\sum_{k=1}^{n} e^{S_{ik}}}.
\end{equation}

In practice, softmax is implemented using a numerically stable formulation that subtracts the maximum logit
\begin{equation}
A_{ij} =
\frac{e^{S_{ij}-m_i}}{\sum_{k=1}^{n} e^{S_{ik}-m_i}},
\qquad
m_i=\max_k S_{ik}.
\end{equation}

Since softmax is invariant to constant shifts of the input, this operation does not change the output but prevents numerical overflow, as discussed in \cite[Ch.~4, p.~81]{Goodfellow-et-al-2016}. The normalization enforces $\sum_i \alpha_i = 1$, producing a probability distribution over tokens and introducing competition between them.

The computational complexity of attention is dominated by the query–key multiplication, resulting in $O(n^2 \cdot d)$. The softmax operation itself has $O(n^2)$ arithmetic complexity, but introduces additional architectural overhead due to row-wise reduction operations (maximum and summation) followed by normalization. These steps create data dependencies across each row, requiring buffering of intermediate results and increasing memory access overhead, which limits fully streaming implementations and increases latency in parallel hardware architectures.

\begin{figure}[tb!]
\centering
\includegraphics[width=1.0\linewidth]{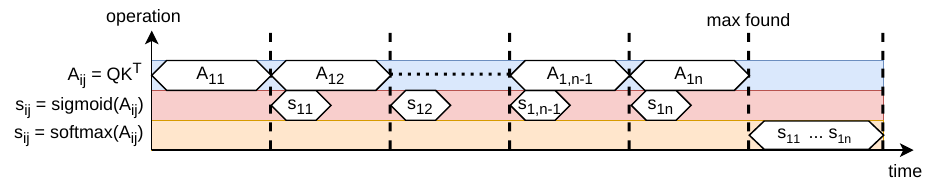}
\caption{Illustration of operation scheduling and dataflow comparison between softmax and sigmoid attention for the first row of the $A_{ij}$ matrix.}
\label{fig:dataflow_softmax_sigmoid}
\end{figure}

\subsection{Sigmoid-Based Self-Attention}

An alternative is to replace softmax with the sigmoid activation, which is applied element-wise to the attention score matrix $\mathbf{S}$:
\begin{equation}
A_{ij} = \sigma(S_{ij}) = \frac{1}{1 + e^{-S_{ij}}}.
\end{equation}
Unlike softmax, sigmoid does not require normalization and produces outputs in the range $(0,1)$, allowing multiple tokens to receive high attention weights concurrently.  

From a computational perspective, sigmoid attention retains the overall $O(n^2 \cdot d)$ complexity due to the matrix multiplications. However, the activation stage has $O(n^2)$ complexity and requires no reduction operations, enabling streaming execution and parallel computation without intermediate buffering. Fig.~\ref{fig:dataflow_softmax_sigmoid} illustrates this dataflow, highlighting the reduction in memory traffic and improvement in parallel hardware utilization compared to softmax. 

\subsubsection{Sigmoid with bias}

A limitation of classical sigmoid attention is that the expected magnitude of the attention weights grows with the sequence length~\cite{softmax_sigmoid2}. If the attention scores are centered around zero, the sigmoid output is approximately $0.5$, which results in a row sum proportional to the number of tokens.
To stabilize the magnitude of the attention weights, a bias term $b$ can be added before the sigmoid activation~\cite{softmax_sigmoid2}:

\begin{equation}
A_{ij} = \sigma(S_{ij}+b), 
\qquad
b=-\log(n),
\end{equation}
where $n$ is the sequence length. This shift reduces the expected sigmoid output such that the total attention weight per row remains approximately constant as $n$ increases, preventing the attention output from growing with sequence length. 
Since the bias introduces only a constant addition per element, the computational complexity remains $O(n^2)$ for the activation stage, preserving the fully parallel, element-wise nature of the computation without additional buffering or control overhead.

\subsubsection{Sigmoid with normalization}

Another alternative is to explicitly normalize the sigmoid outputs across each row:

\begin{equation}
\tilde{A}_{ij} =
\frac{A_{ij}}{\sum_{k=1}^{n} A_{ik}} .
\end{equation}

This operation ensures that each row sums to one, restoring a probabilistic interpretation while retaining the element-wise nature of the sigmoid activation. However, it reintroduces a reduction step requiring row-wise accumulation and buffering before normalization, resulting in partial data dependencies across each row. Compared to softmax, this approach avoids exponential computations, reducing the overall arithmetic complexity of the activation stage and yielding a more hardware-efficient implementation.

The hardware implementation and resulting localization accuracy of these activation functions are described and evaluated in the following section.

\section{Hardware Implementation}

\begin{figure}[tb!]
  \centering 
  \includegraphics[width=1.0\linewidth]{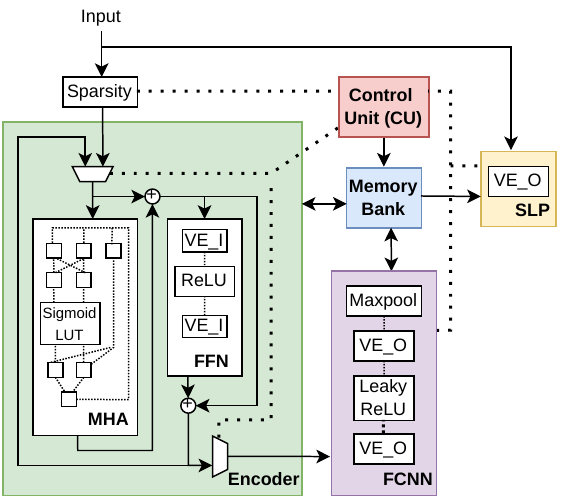}
  \caption{Simplified top-level overview of the hardware architecture.}
  \label{fig:hardware_top_level}
\end{figure}

The top-level architecture of the proposed accelerator is shown in Fig.~\ref{fig:hardware_top_level}. A lightweight SLP first determines the propagation scenario and selects the corresponding weights and biases for the encoder and FCNN layers based on the input features of a single snapshot. The input is then processed by the sparsity detection unit, which identifies skippable rows and generates control signals for computation pruning. A centralized control unit (CU), implemented as a finite state machine, controls and sequences the execution of all submodules, ensuring deterministic dataflow. Memory banks store model parameters and intermediate data in a contiguous layout, enabling efficient reuse across computation stages. Selected weight matrices are stored in transposed form to support column-wise access. The encoder layer, comprising the attention module and the FFN, is reused across multiple layers through folding. This approach enables support for varying network depths using a single hardware instance, reducing resource utilization at the cost of increased latency due to sequential execution. The final localization output is produced by the FCNN module.

The proposed hardware employs multiple vector engines with $23$, $32$, $46$, and $64$ processing elements (PE) tailored to the workload characteristics of different stages. To enhance data reuse, support row-wise sparsity, and reduce data movement overhead, a mixed dataflow strategy is adopted. 
In the input-stationary scheme, activations are retained locally within PEs while weights are streamed, enabling reuse across multiple multiplications (e.g., Q, K, V projections) and reducing repeated memory accesses. 
In contrast, the output-stationary scheme accumulates partial sums locally, minimizing write-back operations. This is used in stages where both operands vary across computations (e.g., single-head attention outputs), or where alternative mappings would significantly increase the required number of multipliers (e.g., SLP). These complementary strategies are assigned per stage and are denoted as VE\_I (input-stationary) and VE\_O (output-stationary) in Fig.~\ref{fig:hardware_top_level}.

Symmetric quantization is adopted to support efficient sparsity-aware hardware implementation. In symmetric schemes, the real value zero maps exactly to the integer value zero, preserving bit-exact zeros throughout the computation and enabling efficient zero detection for row-skipping, whereas asymmetric quantization introduces an offset that complicates sparsity detection. Activations and weights are represented using 16-bit signed integers in Q8.8 format, while multiply-accumulate operations are performed in extended precision to preserve dynamic range during accumulation. Further reduction of bit width is left for future work.

\subsection{Adaptive Model Selection and Sparsity Support}

\begin{figure}[tb!]
  \centering 
  \includegraphics[width=0.9\linewidth]{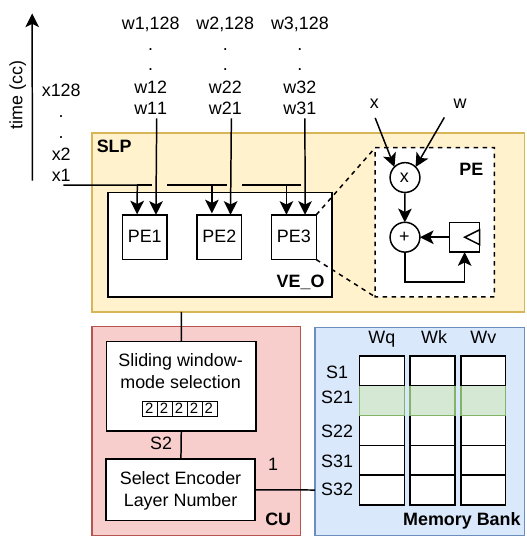}
\caption{Simplified block diagram of adaptive model selection using the SLP, CU, and memory bank. Only the logic for selecting model-specific weights and biases is shown. The SLP is mapped to an output-stationary PE array with the corresponding input dataflow illustrated. The first and second encoder layers of S2 and S3 are denoted as S21, S22, S31, and S32, respectively.}
  \label{fig:hardware_SLP}
\end{figure}

To improve the robustness of the adaptive model selection, a sliding-window mode selection mechanism is applied to the router outputs, as shown in Fig.~\ref{fig:hardware_SLP}. The predicted scenario labels for each snapshot are buffered over a fixed-length window, and the selected model corresponds to the majority-vote class within the window. This suppresses transient misclassifications and stabilizes switching between specialized models. During scenario transitions, the window has a transition period, as previously buffered labels are gradually replaced. The memory bank stores a concatenation of weights and biases from all scenarios and their corresponding encoder layers, partitioned into five segments (S1, S21, S22, S31, and S32). The CU selects the appropriate segment based on the router output and execution stage.

\begin{figure*}[tb!]
\centering
\includegraphics[width=1.0\linewidth]{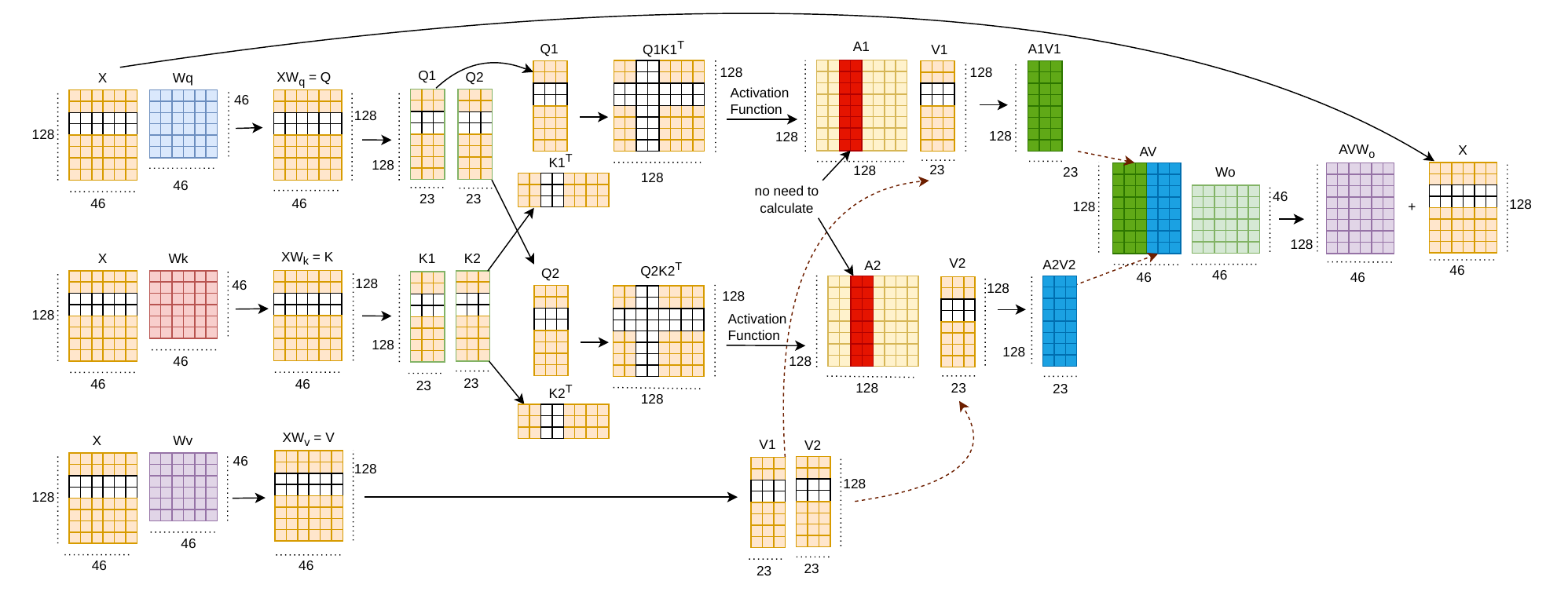}
\caption{Row-skipping mechanism after element-wise thresholding, where a zero-count criterion $(Z_i > T_r)$ generates a mask that gates PE activity and memory accesses. Skipped rows are shown in white, corresponding to rows removed from subsequent vector computations.}
\label{fig:row_skip_logic}
\end{figure*}

The proposed hardware architecture exploits sparsity in the beam-delay channel representation by identifying rows with limited contribution and skipping their computations. First, element-wise thresholding suppresses low-magnitude components, followed by row-wise sparsity detection to identify rows containing predominantly near-zero elements. Lightweight control logic generates a row mask that disables the corresponding memory accesses, allowing entire rows to be skipped during vector computations. As illustrated in Fig.~\ref{fig:row_skip_logic}, this reduces the number of active rows processed in the Q, K, V projections and subsequent attention computation. Operating at a coarse granularity, the mechanism incurs minimal control overhead while enabling significant reductions in redundant computations. The design prioritizes flexible, sparsity-aware row-wise processing over systolic arrays or fixed matrix-multiplication engines and supports variable effective sequence lengths through data-dependent row skipping.

\subsection{Encoder Layer} 

\begin{figure}[tb!]
\centering
\includegraphics[width=1.0\linewidth]{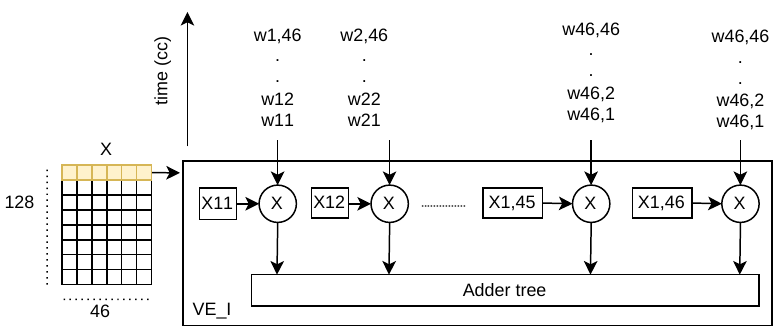}
\caption{Microarchitecture of the input-stationary vector engine with 46 multipliers used in MHA.}
\label{fig:pe_microarch}
\end{figure}

The MHA stage, which constitutes approximately $80\%$ of the total execution time (for a single encoder layer), is optimized by differentiating dataflow strategies between the projection and core attention phases. The Q, K, and V projections are implemented using a vector engine composed of parallel PEs and an adder tree, as illustrated in Fig.~\ref{fig:pe_microarch}. These stages follow a row-buffered streaming dot-product pipeline with input-stationary kernels, where each input row is retained locally and reused across streamed weight vectors. The adder tree aggregates partial sums to produce the projection outputs.

The $\mathbf{Q}\mathbf{K}^T$ score computation is implemented as a streaming operation without stationary accumulation. In contrast, the subsequent $\phi(\mathbf{Q}\mathbf{K}^T)\mathbf{V}$ stage uses output-stationary accumulation, where partial products are accumulated locally within the PEs for each output row. The implementation of the activation function is described in the next subsection.

The FFN, consisting of two layers, follows input-stationary kernels with row-buffer reuse, enabling efficient processing of intermediate activations. Overall, these mappings ensure efficient execution, with the FFN contributing approximately $15\%$ of the total processing time.

\subsection{Activation Function}

The choice of activation function directly impacts both hardware complexity and localization performance. In this subsection, different activation functions are evaluated using both floating-point and integer-only models. Integer-only softmax attention is implemented as in SwiftTron~\cite{swifttron}. The softmax function requires maximum reduction, exponentiation, accumulation, and division, introducing row-wise data dependencies and limiting parallelism. In contrast, sigmoid attention enables fully element-wise computation and is realized using a lookup table (LUT), eliminating the need for exponentiation and global normalization, and allowing a single-pass, streaming implementation with reduced latency and control overhead.

The sigmoid function is implemented using a LUT with $1025$ uniformly spaced samples over $[-16,+16]$ at a resolution of $1/32$, resulting in a memory requirement of approximately $16.4$\,kbits ($\approx 2$\,kB). The biased sigmoid variant introduces only a constant offset prior to activation and incurs negligible hardware overhead, while the normalized sigmoid reintroduces row-wise accumulation and reciprocal operations but remains less complex than softmax due to the absence of exponential computations.

Table~\ref{tab:summary_activation_functions} summarizes the localization accuracy across scenarios for both floating-point and integer-only models. The results demonstrate that LUT-based sigmoid attention maintains competitive performance while reducing computational and implementation cost compared to softmax attention. Among the evaluated variants, the biased sigmoid provides the most favorable trade-off between accuracy and hardware efficiency and is therefore selected in the final design.

\begin{table}[htbp!] 
\caption{Mean error (ME) in meters for different activation functions (10 trials). Integer-only results correspond to hardware-mapped implementations.}
\begin{center}
\begin{tabular}{|l|ccc|ccc|}
\hline
\textbf{Activation} & \multicolumn{3}{c|}{\textbf{Float}} & \multicolumn{3}{c|}{\textbf{Integer Only}} \\ \cline{2-7}
 & S1 & S2 & S3 & S1 & S2 & S3 \\ \hline
reference~\cite{yaman2025adaptiveattentionbasedmodel5g} & 0.4 & 0.78 & 0.77 & -- & -- & -- \\ \hline
softmax & 0.39 & 1.0 & 0.68 & 0.43 & 1.02 & 0.76 \\ \hline
sigmoid & 0.53 & 1.25 & 1.16 & 0.54 & 1.21 & 1.23 \\ \hline
sigmoid w. bias & 0.38 & 0.82 & 0.74 & \textbf{0.51} & \textbf{0.98} & \textbf{0.83} \\ \hline
sigmoid w. norm & 0.4 & 0.76 & 0.61 & 0.43 & 0.89 & 0.66 \\ \hline
\end{tabular}
\label{tab:summary_activation_functions}
\end{center}
\end{table}

\subsection{FCNN}

The final FCNN stage is preceded by a Max-Pooling layer, which significantly reduces the spatial dimensions of the feature map. Given the reduced data volume and the requirement for a final localization output vector, the FCNN is mapped to an output-stationary dataflow. This ensures that the accumulation of the final coordinate outputs remains localized within the PE array. By tailoring the stationary strategy to the specific mathematical characteristics of each layer, the architecture achieves efficient execution, with the FCNN contributing only approximately $5\%$ to the total inference time.

\section{Results}
This section evaluates the localization accuracy and hardware characteristics of the proposed adaptive architecture across three propagation scenarios (S1, S2, and S3). The design is implemented on the Xilinx Zynq UltraScale+ ZCU104 platform using Q8.8 fixed-point arithmetic. In this table and implementation, the sparsity threshold for S3 is selected as $T = 0.006$ with a corresponding zero-count threshold of $28$. This choice is constrained by the quantization resolution, as the minimum representable value with 8-bit fractional precision is approximately $0.0039$, and by the requirement to maximize row sparsity while limiting the ME degradation to below $10\%$. Among the evaluated configurations, this setting provides the highest achievable sparsity under the specified accuracy constraint. Under this configuration, a row sparsity of $6.98\%$ is achieved with an ME degradation of $9.4\%$. 

To quantify the contribution of each design component, an ablation study is performed by progressively introducing sigmoid attention, quantization, and sparsity. This analysis isolates the impact of each optimization on localization accuracy and highlights the trade-offs between numerical precision and computational efficiency.
The resulting localization performance, including both mean error (ME) and the $95\%$ confidence level (CL), is summarized in Table~\ref{tab:ablation_studies}. 
The $95\%$ CL metric is particularly relevant for practical positioning systems, as 3GPP positioning requirements are commonly specified in terms of confidence levels, including sub-meter accuracy targets for selected use cases~\cite{3GPP_TS_22261, 3GPP_TR_38857}.

\begin{table}[htbp!]
\caption{Ablation study of optimization techniques. Mean error (ME) and 95\% confidence level (CL) in meters across scenarios.}
\centering
\begin{tabular}{|l|ccc|ccc|}
\hline
\textbf{Method} & \multicolumn{3}{c|}{\textbf{ME (m)}} & \multicolumn{3}{c|}{\textbf{CL (m)}} \\ \cline{2-7}
 & S1 & S2 & S3 & S1 & S2 & S3 \\ \hline
Reference~\cite{yaman2025adaptiveattentionbasedmodel5g} & 0.40 & 0.78 & 0.77 & -- & -- & -- \\ \hline
+ Sigmoid w. bias & 0.38 & 0.82 & 0.74 & 0.77 & 1.85 & 1.63 \\ \hline
+ Quantization & 0.51 & 0.98 & 0.83 & 0.99 & 2.23 & 1.75 \\ \hline
+ Sparsity & 0.51 & 1.13 & 0.88 & 0.99 & 2.50 & 1.86 \\ \hline
\end{tabular}
\label{tab:ablation_studies}
\end{table}

The average row sparsity is $9.5\%$, $13.9\%$, and $6.92\%$ for S1, S2, and S3, respectively, while the maximum number of skipped rows per snapshot reaches $65\%$, $59\%$, and $58\%$. This yields computational speedups of up to \textbf{$2.08\times$}, \textbf{$1.32\times$}, and \textbf{$1.31\times$}, reducing the inference latency from $1.06$\,ms, $2.11$\,ms, and $2.11$\,ms to $0.51$\,ms, $1.60$\,ms, and $1.61$\,ms for S1, S2, and S3, respectively. The highest speedup is observed for the single-layer encoder configuration in S1, which is expected since only input sparsity is exploited, directly affecting the first encoder layer. In contrast, for S2 and S3, the second encoder layer dominates the overall latency, thereby limiting the achievable speedup.

While hardware accelerators for Transformers (e.g., SwiftTron, SpAtten) and 5G localization (e.g., CNN or MUSIC-based) exist independently, this work proposes an adaptive shallow Transformer architecture specifically optimized for the sparsity and latency characteristics of 5G radio-based localization.
%this work is the first to propose an Adaptive Shallow Transformer architecture specifically optimized for the unique sparsity and latency requirements of 5G radio signals. 
Table~\ref{tab:comparison_table} summarizes the hardware performance relative to a CPU baseline and an ASIC implementation of CNN-based localization using massive MIMO~\cite{Mohammad_mimo_pos}. Despite operating at a significantly lower clock frequency of \SI{100}{\mega\hertz}, 
the FPGA implementation achieves a throughput of up to $1961$\,pos/s for the single-layer S1 scenario.  
Even in the multi-layer S2 and S3 configurations, our implementation achieves a higher reported positioning rate than the $555$\,MHz ASIC. However, a direct comparison is limited by differences in algorithm, implementation technology, dataset, and operating environment. 
While the ASIC implementation in \cite{Mohammad_mimo_pos} reports a lower root mean square error (RMSE), it is evaluated in a small-scale indoor environment, whereas our design achieves $(0.58, 1.35, 1.04)$,m RMSE across the outdoor S1, S2, and S3 scenarios.

\begin{table}[htbp!]
\caption{Hardware Performance Across Platforms}
\centering
\begin{tabular}{|l|c|c|c|}
\hline
\textbf{Metric} & \textbf{\cite{yaman2025adaptiveattentionbasedmodel5g}} & \textbf{\cite{Mohammad_mimo_pos}} & \textbf{This Work}  \\ \hline
    Architecture & CPU & 22nm ASIC & FPGA \\ \hline
    Clock Freq. (MHz) & 3600 & 555 & 100 \\\hline
    Algorithm & Transformer & CNN & Transformer \\\hline
    Environment & Outdoor & Indoor$^{\mathrm{a}}$ & Outdoor \\\hline
    Arithmetic & FP32 & INT16 & INT16 \\\hline
    RMSE (m) & 0.50$^{\mathrm{b}}$ & 0.4 & 0.58$^{\mathrm{b}}$ \\\hline 
    Inference Time (ms) & 3.97-4.10$^{\mathrm{c}}$ & 3.69 & 0.51-2.11 \\\hline
    Peak Throughput & 252 & 271 & 1961 \\
    (pos/s) &   &  &  \\\hline
    Total Power (W) & N/A & 0.15 & 1.29 \\ \hline
    Energy per  & N/A & 0.55 & 0.66-2.74 \\
    Inference (mJ) &&& \\ \hline
\multicolumn{4}{l}{$^{\mathrm{a}}$Based on simulated data.} \\ 
\multicolumn{4}{l}{$^{\mathrm{b}}$Average over single-snapshot evaluation in S1.} \\  
\multicolumn{4}{l}{$^{\mathrm{c}}$Per-snapshot inference time; model loading excluded.} \\ % with loading model 8ms, without loading model:4ms
\end{tabular}
\label{tab:comparison_table}
\end{table}

Utilization results are given in Table~\ref{tab:fpga_results}. Vivado power analysis estimates a total on-chip power consumption of approximately $1.29$\,W, with $0.702$\,W attributed to dynamic power and $0.597$\,W to device static power. Among the dynamic components, signals and logic contribute the largest share, while BRAM and DSP blocks account for approximately $22\%$ and $24\%$ of the dynamic power respectively. 

\begin{table}[htbp!]
\caption{FPGA implementation results on Zynq UltraScale+ ZCU104.}
\centering
\begin{tabular}{|l|c|c|c|}
\hline
\textbf{Resource} & \textbf{Used} & \textbf{Available} & \textbf{Utilization} \\ \hline
CLB LUT & 33296 & 230400 & 14.4\% \\ \hline
LUTRAM & 812 & 101760 & 0.8\% \\ \hline
FF & 26281 & 460800 & 5.7\% \\ \hline
BRAM & 196.5 & 312 & 63.0\% \\ \hline
DSP & 529 & 1728 & 30.6\% \\ \hline
I/O & 36 & 360 & 10.0\% \\ \hline
\end{tabular}
\label{tab:fpga_results}
\end{table}

\section{Conclusion}
A hardware-efficient, sparsity-aware Transformer accelerator is developed for outdoor localization in a massive MIMO infrastructure. By exploiting structured sparsity of beam-delay channel representations, the proposed architecture implements a row-wise skipping mechanism that dynamically bypasses redundant computations with minimal control overhead. To mitigate the hardware bottleneck of the attention layer, the globally dependent softmax activation is replaced with localized, hardware-friendly sigmoid-based variants. We show that biased sigmoid attention allows for a single-pass streaming dataflow, eliminating the synchronization barriers and complex divider logic required by traditional softmax. The design was implemented on an FPGA using a heterogeneous vector processing array and symmetric fixed-point quantization. Experimental results across diverse propagation scenarios demonstrate that the accelerator achieves a mean error of less than $1.15$,m while maintaining high throughput. 
Depending on the environmental complexity and model configuration, the accelerator achieves an inference latency of $0.51$-$2.11$,ms, meeting the considered sub-10-ms latency target.
\bibliographystyle{ieeetr} 
\bibliography{main}

\section*{} % Empty header, to properly separate references from biographies.
\vskip -4\baselineskip plus -1fil
\begin{IEEEbiography}[{\includegraphics[width=1in,height=1.25in,clip,keepaspectratio]{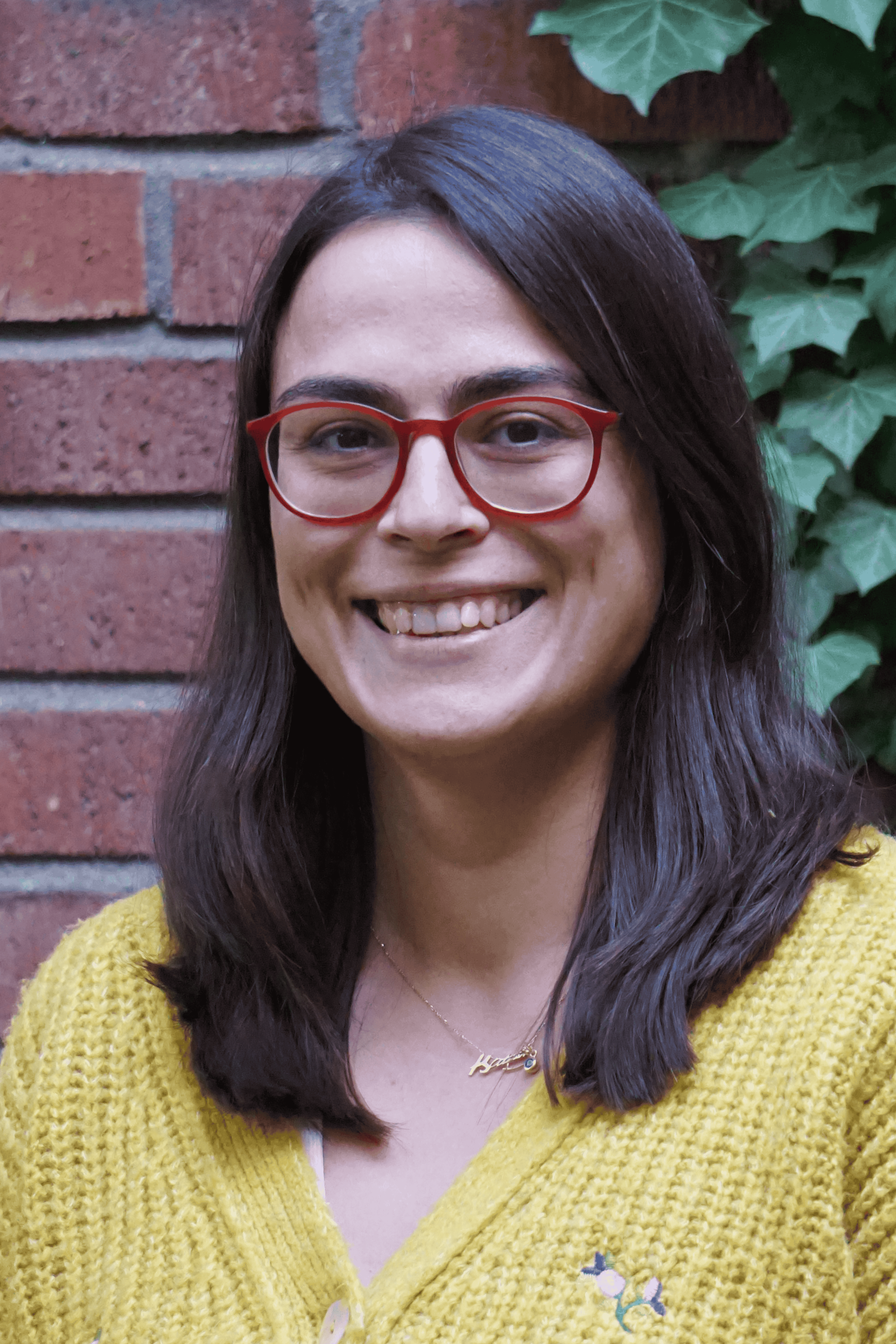}}]
 {Ilayda Yaman}~(Student Member, IEEE) 
completed her bachelor’s degree at Istanbul Technical University in 2018 and her master’s degree in Embedded Electronics Engineering at Lund University in 2020. During her master's degree, she received the LU Global Scholarship. Currently, she is a Ph.D. student at Lund University (Main Supervisor: Liang Liu). Her current research area is low-power ML hardware for radio-based localization systems.
\end{IEEEbiography}

\vskip -1\baselineskip plus -1fil
\begin{IEEEbiography}[{\includegraphics[width=1in,height=1.25in,clip,keepaspectratio]{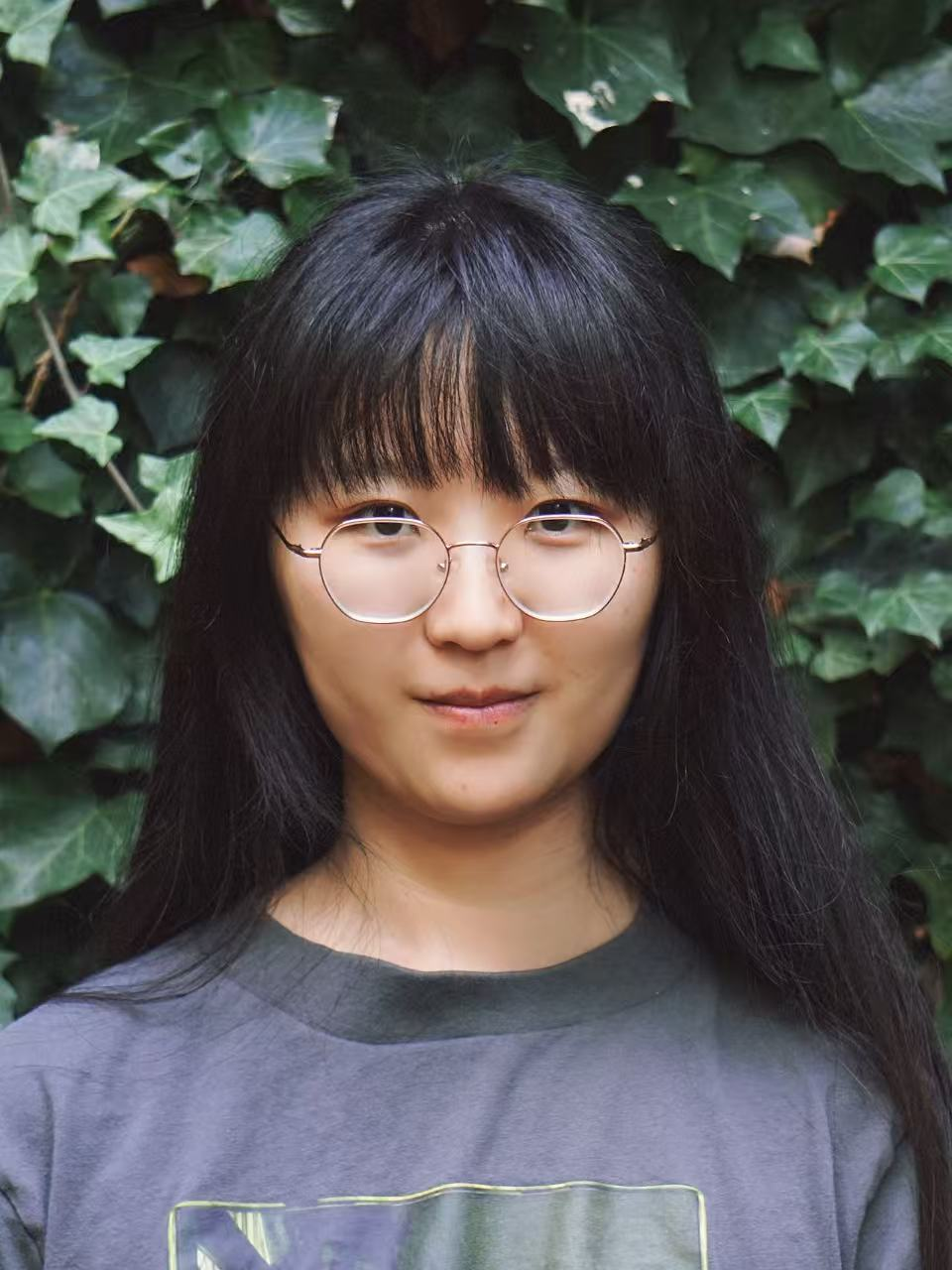}}]{Sijia Cheng}~(Student Member, IEEE) 
received a double bachelor’s degree from Beijing Jiaotong University and KU Leuven in 2020, followed by a master’s degree in Embedded Electronics Engineering from Lund University in 2022.  During her studies at Lund, she received the LU Global Scholarship. She is currently working toward the Ph.D. degree at Lund University under the supervision of Liang Liu. Her research focuses on low-latency digital signal processing for massive MIMO systems.\end{IEEEbiography}

\vskip -1\baselineskip plus -1fil
\begin{IEEEbiography}[{\includegraphics[width=1in,height=1.25in,clip,keepaspectratio]{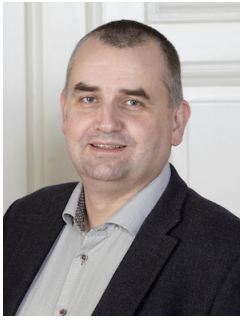}}] {Ove Edfors}~(Senior Member, IEEE) is Professor of Radio Systems at the Department of Electrical and Information Technology, Lund University, Sweden. His research interests include statistical signal processing and low complexity algorithms with applications in wireless communications. In the context of Massive MIMO and large intelligent surfaces, his main research focus is on how realistic propagation characteristics influence system performance and base-band processing complexity
 \end{IEEEbiography}

\vskip -1\baselineskip plus -1fil
\begin{IEEEbiography}[{\includegraphics[width=1in,height=1.25in,clip,keepaspectratio]{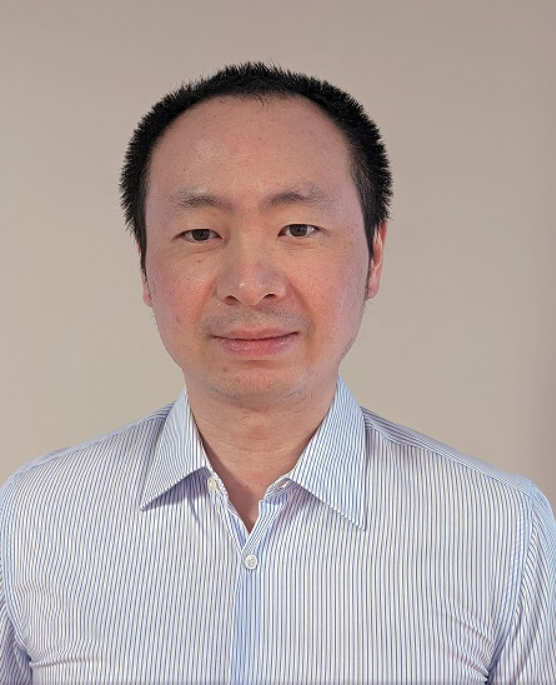}}] {Liang Liu}~(Member, IEEE) received his Ph.D. degree from Fudan University in 2010. He joined Lund University as post-doc. Since 2024, he has been a Professor at Lund University. His research interests include wireless systems and digital integrated circuit design. He is a member of the Technical Committee of VLSI Systems and Applications and CAS for Communications of the IEEE Circuit and Systems Society.
 \end{IEEEbiography}

\end{document}